\documentclass[%
reprint,
superscriptaddress,
amsmath,amssymb,
aps,
]{revtex4-2}
\usepackage{easyReview}
\usepackage{graphicx}
\usepackage{dcolumn}
\usepackage{bm}
\usepackage{hyperref}
\usepackage{soul}
\usepackage{braket}

\usepackage{xcolor}
\usepackage{physics}
\usepackage{ulem}

\newcommand{\rt}{\mathrm{t}}
\newcommand{\rr}{\mathrm{r}}
\newcommand{\rb}{\mathrm{b}}
\newcommand{\vp}{\mathcal{A}}
\newcommand{\lnk}[1]{\langle #1 \rangle}

\begin{document}
\title{Higher-dimensional Hofstadter butterfly on Penrose lattice}
\author{Rasoul Ghadimi}
\affiliation{Center for Correlated Electron Systems, Institute for Basic Science (IBS), Seoul 08826, Korea}
\affiliation{Department of Physics and Astronomy, Seoul National University, Seoul 08826, Korea}
\affiliation{Center for Theoretical Physics (CTP), Seoul National University, Seoul 08826, Korea}

\author{Takanori Sugimoto}
\affiliation{Center for Qunatum Information and Qunatum Biology, Osaka University, Osaka 560-0043, Japan}
\affiliation{Advanced Science Research Center, Japan Atomic Energy Agency, Tokai, Ibaraki 319-1195, Japan}

\author{Takami Tohyama}
\affiliation{Department of Applied Physics, Tokyo University of Science, Tokyo 125-8585, Japan}
\date{\today}
\begin{abstract}
Quasicrystal is now open to search for novel topological phenomena enhanced by its peculiar structure characterized by an irrational number and high-dimensional primitive vectors.
Here we extend the concept of a topological insulator with an emerging staggered local magnetic flux (i.e., without external fields), similar to the Haldane's honeycomb model, to the Penrose lattice as a quasicrystal.
The Penrose lattice consists of two different tiles, where the ratio of the numbers of tiles corresponds to an irrational number.
Contrary to periodic lattices, the periodicity of energy spectrum with respect to the magnetic flux no longer exists reflecting the irrational number in the Penrose lattice.
Calculating the Bott index as a topological invariant, we find topological phases appearing in a fractal energy spectrum like the Hofstadter butterfly.
More intriguingly, by folding the one-dimensional aperiodic magnetic flux into a two-dimensional periodic flux space, the fractal structure of energy spectrum is extended to higher dimension, whose section corresponds to the Hofstadter butterfly.
\end{abstract}
\maketitle

Essential characteristics of quasicrystalline physical properties have continuously been sought and discussed since the astonishing discovery of quasicrystals~\cite{Shechtman1984}, because of their distinct characteristics; higher-order (5, 8, or 10-fold) rotational symmetry, an irrational ratio of the numbers of different local structures, fractality in their global structure, and the higher-dimensional primitive vectors, instead of the translational symmetry~\cite{Shechtman1984,Levine1986,Socolar1986}.
The first investigations on single-particle electronic properties have been performed more than three decades ago~\cite{Arai1988,Fujiwara1988,Tokihiro1988,Ma1989,Fujiwara1991,Liu1991,Tsunetsugu1991,Koga2017,Koga2020,Koga2022}, resulting in several discoveries of quantum properties in quasicrystals, e.g., (critical or confined) zero-energy eigenstates~\cite{Liu1991,Tsunetsugu1991,Koga2017,Koga2020,Koga2022} and (multi-)fractal structures in energy spectrum~\cite{Tokihiro1988,Yamamoto1995,Hatakeyama1998,Bandres2016}.
In parallel, thermodynamical properties in quasicrystals have also retained much interest, because of a specific lattice degree of freedom, the so-called phason, corresponding to hidden degree of freedom related to the higher dimension~\cite{Yamamoto1995,Szallas2009,[{As a review, see }]Janssen2018}.
Furthermore, recent experimental discoveries of quasicrystalline ferromagnetism, superconductivity, and quantum criticality~\cite{Tamura2021,Kamiya2018,Deguchi2012,[{As a review, see }]Sato2022}, and successful realizations of aperiodic optical lattices~\cite{Guidoni1997,Bandres2016} and topological photonic quasicrystals~\cite{Kraus2012,Kraus2013,Verbin2013} have spurred theoretical investigations of exotic physical properties in quasicrystals.
In addition, recent attempts to discover novel phases of matter, have been focused on topological phases in quasicrystals; topological insulators~\cite{Tran2015,Fuchs2016,Huang2018,Huang2018-2,Huang2019,Fuchs2018,He2019,Chen2019,Spurrier2020,Chen2020,Duncan2020}, topological superconductors~\cite{Ghadimi2017,Varjas2019,Cao2020,Ghadimi2021}, higher-order topological phases~\cite{Chen2020,Peng2021}, and hidden topologies in non-Hermitian systems~\cite{Longhi2019,Weidemann2022}.

Despite the recent progress in topological phases of quasicrystals, essential properties in neither crystals nor amorphous systems but quasicrystals have almost not been clarified so far.
To extract an essential property common in the quasicrystals, we focus on the irrational number characterizing the quasicrystalline structure.
In the quasicrystals, the irrational number corresponds to the ratio of the numbers of two different tiles and the ratio of surfaces of the tiles, e.g., the golden ratio $(1+\sqrt{5})/2(=\tau)$ for the Penrose lattice [see Fig.~1(a)] and the silver ratio $1+\sqrt{2}$ for the Ammann-Beenker lattice~\cite{Senechal1996,Walter2009}.
Yet, how do we grab an advent of the irrational number in physical quantity?
Here, we propose a model of a quasicrystalline topological insulator, which is similar to that used by C. W. Duncan {\it et al.}~\cite{Duncan2020} but newly includes an emerging staggered local magnetic flux (i.e., no external fields). Our proposed model, thus, corresponds to an extension of the Haldane's honeycomb model~\cite{Haldane1988} to quasicrystals.

In the Duncans' work~\cite{Duncan2020}, they apply a uniform magnetic field and assume local magnetic fluxes proportional to the surfaces of two tiles.
Since the ratio of surfaces corresponds to the irrational number in quasicrystals, the ratio of local fluxes becomes the irrational number and breaks the periodicity of energy spectrum including topological phases with respect to the magnetic flux.
However, as mentioned in \cite{Duncan2020}, the irrational ratio of surfaces is realized even in uniform crystals.
On the other hand, the irrational ratio of the numbers of two tiles is undoubtedly unique to quasicrystals.
Instead of a uniform magnetic field, if we apply a totally-zero staggered magnetic field and impose an equivalent local magnetic flux on the same type of tiles, the ratio of the fluxes can be the irrational number via the ratio of the numbers of tiles.
In fact, with the zero external field condition $N_\rr\phi_\rr+N_\rb\phi_\rb=0$, the ratio of local fluxes is equivalent to the ratio of the numbers of tiles, i.e., the irrational number, $\phi_\rr/\phi_\rb=-N_\rb/N_\rr$, where $N_\rr (N_\rb)$ and $\phi_\rr (\phi_\rb)$ represents the number of red (blue) tiles in Fig.~1(a) and the local flux of them, respectively.
Therefore, with a staggered magnetic field, the broken periodicity of energy spectrum appears only in quasicrystals.
Moreover, the aperiodic structure with respect to one of magnetic fluxes can be folded into two-dimensional periodic flux space.
Adding an energy axis to the two-dimensional flux space, we can see a three-dimensional complicated structure of energy spectrum, whose vertical section is the so-called Hofstadter butterfly.

\begin{figure}[t!]
	\centering
				\includegraphics[width=\linewidth]{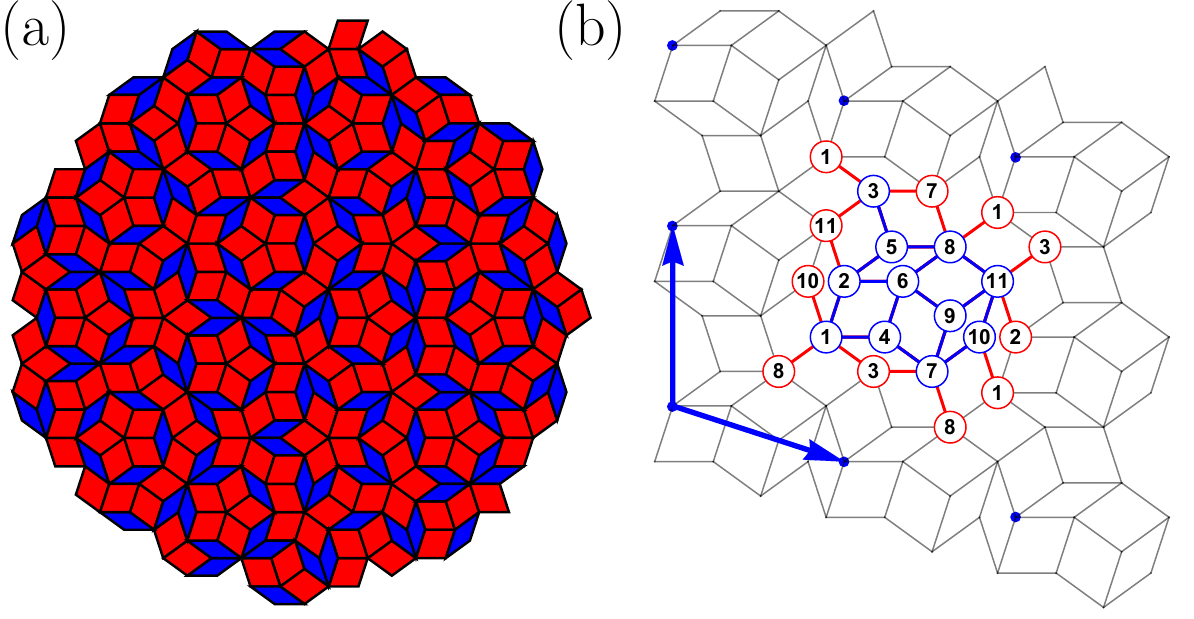}
	\caption{(a) Penrose lattice constructed by fat (red) and thin (blue) tiles.
		(b) A $g=1$ approximant of the Penrose lattice, where the 1st to 11th blue sites are linked by blue bonds.
        Thanks to the approximant periodicity, we regard the lattice as a unit cell.
				The red edges linking blue to red sites denote the bonds bridging between neighboring unit cells, corresponding to connection of boundaries with the periodic boundary condition.
        }
	\label{fig:crystall}
\end{figure}

In the following, as a case study showing this behavior, we focus on the Penrose lattice, though we have also obtained a qualitatively similar result in the Ammann-Beenker lattice (not shown).
The Penrose lattice is a typical example of two-dimensional quasicrystals and is composed of two (fat and thin) rhombuses [see red and blue tiles in Fig.~\ref{fig:crystall}(a)].
Although translational symmetry is absent, it has various structural properties, such as five-fold symmetry and self-similarity related to the inflation/deflation rules~\cite{Senechal1996,Walter2009}.
To demonstrate numerical results, we use the approximant method to generate Penrose quasicrystal~\cite{Tsunetsugu1991,Ghadimi2020}.
The quasicrystal approximant has a translationally symmetric structure with a unit cell resembling a local structure of the original quasicrystal [see Fig.~\ref{fig:crystall}(b)].
The unit-cell size is increased with increasing the approximant generation $g$, so that real quasicrystal is obtained with $g\to\infty$.
The numbers of the red and blue tiles in $g$th generation are $N_\rr^{(g)}=3F_{2g+1}+F_{2g}$ and $N_\rb^{(g)}=F_{2g+1}+2F_{2g}$, respectively, with the $i$th Fibonacci number $F_{i}$.
According to the inflation rule between $g$th and $(g-1)$th generations, the number of tiles increases according to, $\bm{N}_{g}=\mathbf{F}\bm{N}_{g-1}$ with
\begin{equation}
	\bm{N}_{g} = \begin{pmatrix} N_{\rr}^{(g)}& N_\rb^{(g)} \end{pmatrix}^T, \
	\mathbf{F}= \begin{pmatrix} 2 & 1 \\ 1 & 1\end{pmatrix}.
\end{equation}
Since the eigenvalues of the inflation matrix $\mathbf{F}$ are $\{(1\pm\sqrt{5})/2\}^2$, in the thermodynamic limit $g\to\infty$, the component of the number's vector $\bm{N}_g$ along the eigenvector with the smaller eigenvalue vanishes.
The eigenvector for the larger eigenvalue is $(\tau,1)$ with the golden ratio $\tau$, so that the ratio of the numbers of two tiles converges to the golden ratio, $N_\rr^{(g)}/N_\rb^{(g)}\to\tau$ [see Table~I].
This is a feature peculiar to quasicrystals, while in the periodic lattices consisting of several types of plaquettes, the ratio of the numbers of plaquettes should be rational.
We omit the superscript about generation ${}^{(g)}$ in the following.

\begin{table}[t]
	\centering
	\caption{The numbers of red and blue tiles, the ratio of them, and difference from the golden ratio in the Penrose approximants.}
	\label{table:table}
	\begin{tabular}{c|cccccc}
		& $g$=1 & $g$=2 & $g$=3 & $g$=4 & $g$=5 & $g$=6 \\
		\hline
		$N_r$ & 7 & 18 & 47 & 123 & 322 & 843 \\
		$N_b$ & 4 & 11 & 29 & 76 & 199 & 521 \\
		$\frac{N_r/N_b-\tau}{\tau}$ & $\mathcal{O}(10^{-1})$  &$\mathcal{O}(10^{-2})$ &$\mathcal{O}(10^{-3})$ &$\mathcal{O}(10^{-4})$&$\mathcal{O}(10^{-5})$ &$\mathcal{O}(10^{-6})$ \\
	\end{tabular}
\end{table}

The model Hamiltonian is given by,
\begin{equation}\label{hamil}
\mathcal{H}=- t\sum_{\lnk{i,j}}e^{\imath \vp_{ij}} c_i^{\dagger} c_j + \mathrm{H.c.} -\mu\sum_i c_i^{\dagger} c_i,
\end{equation}
where $c_i^{\dagger}$ ($c_i$) is creation (annihilation) operator of spinless fermion at $i$th vertex on the Penrose lattice, and $\imath$ is the imaginary unit.
The chemical potential and the hopping integral are denoted by $\mu$ and $t$, respectively.
To introduce a flux in a tile, we use the Peierls phase $\vp_{ij}$, corresponding to a line integral of vector potential on the edge $\lnk{i,j}$ from $i$th to $j$th vertices.
In addition,
to obtain a periodic boundary condition, we use one unit cell of  $g$th Penrose approximant [see Fig.~\ref{fig:crystall}(b)].

In the continuum limit, the vector potential $\mathcal{A}(r)$ is related to magnetic flux $\phi_{S}$ penetrating surface $S$ via
\begin{equation}\label{loop}
\phi_S=\oint_{\partial S} \mathcal{A}\cdot d\bm{r},
\end{equation}
where $\partial S$ is the boundary of the surface $S$.
In lattice models, we can use an alternative of (\ref{loop}) given by
\begin{equation}\label{AtoPhi}
\phi_{(i,j,k,l)}=\mathcal{A}_{ij}+\mathcal{A}_{jk}+\mathcal{A}_{kl}+\mathcal{A}_{li},
\end{equation}
where $\phi_{(i,j,k,l)}$ is the magnetic flux passing through a tile constructed by four vertices, $(i, j, k, l)$, numbered counterclockwise.
As mentioned above, we assume
the same value of flux for each type of tiles, where there are
two types of flux $\phi_{(i,j,k,l)}=\phi_\rr, \phi_\rb$ for the red and blue tiles in Fig.~\ref{fig:crystall}(a), respectively.
Moreover, we introduce a constraint on the
flux satisfying totally zero magnetic field $N_\rr\phi_\rr=-N_\rb\phi_\rb$.
In this sense, the two types of flux depend on each other, and are rewritten by
\begin{equation}\label{stag}
	\phi_\rr= 2\pi \frac{n_\rr}{N_\rr},\ \phi_\rb= 2\pi \frac{n_\rb}{N_\rb},
\end{equation}
with a single parameter $n=n_\rr=-n_\rb$ as the normalized flux.
Note that each magnetic flux has periodicity in the phase space of $[0,2\pi]$, i.e., the period for $n_\rr$ ($n_\rb$) corresponds to $N_\rr$ ($N_\rb$).

The Peierls phase $\vp_{ij}$ is determined as follows.
With a fixed normalized flux $n$, we obtain fluxes for two tiles $(\phi_\rr, \phi_\rb)$.
We computationally assign the local Peierls phase $\vp_{ij}$ for edges one by one, satisfying Eq.~(\ref{AtoPhi}).
Note that the configuration of $\vp_{ij}$ is not uniquely assigned, due to the presence of the gauge degree of freedom.
However, physical properties should not depend on the the gauge transformation: $c_i^{\dag}\rightarrow e^{\imath \theta_i} c_i^{\dagger},\ c_i\rightarrow e^{-\imath \theta_i}c_i$ with $\vp_{ij}\rightarrow \vp_{ij}-(\theta_i-\theta_j)$.

\begin{figure}[t!]
	\centering
				\includegraphics[width=\linewidth]{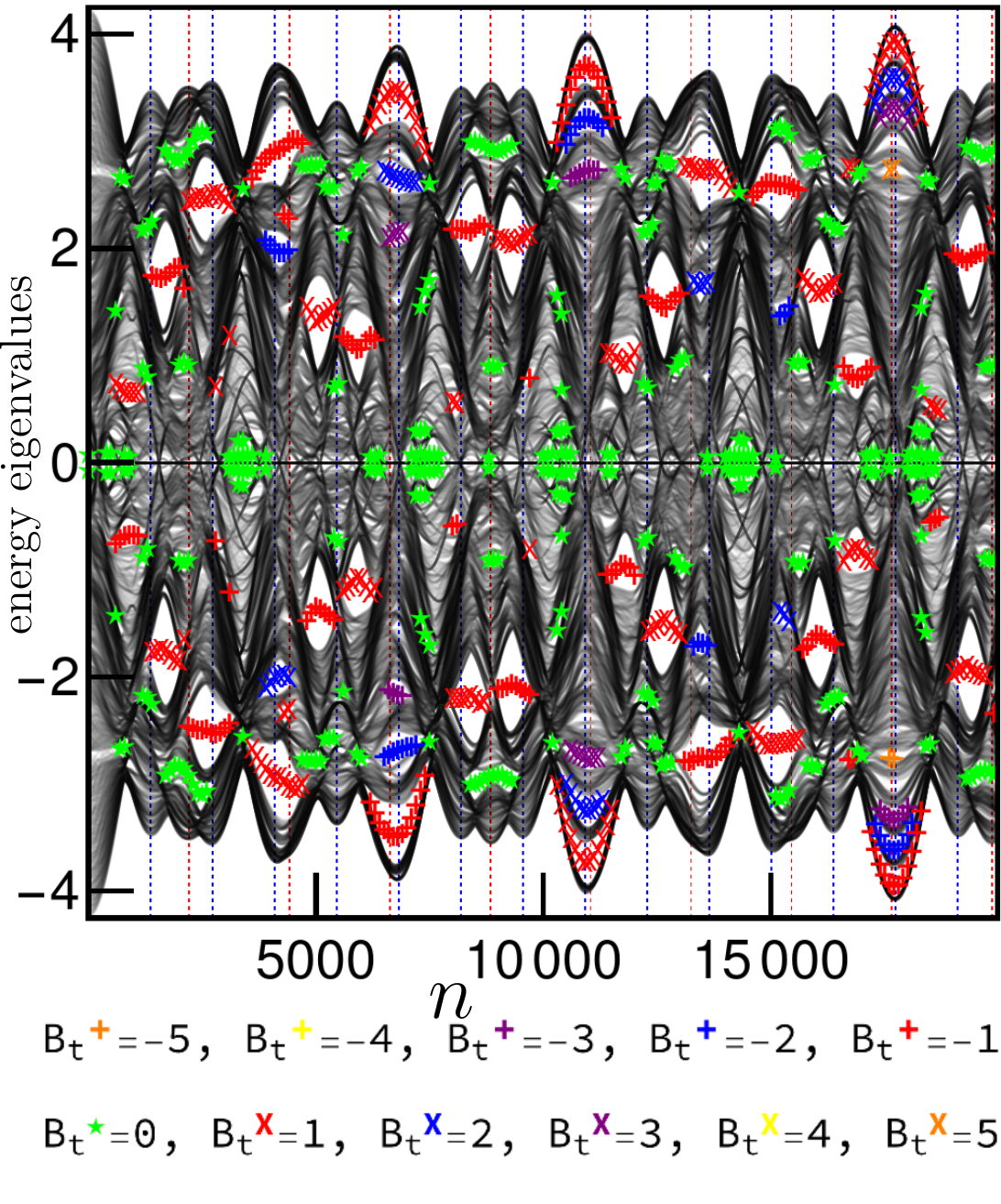}
	\caption{Energy spectrum of a $g=6$ Penrose lattice ($N_\rr=843$ and $N_\rb=521$) with normalized staggered magnetic flux $n$.
          we set the hopping integral $t=1$ as an energy unit.
          The position of colored symbols denotes the chemical potential $\mu$ (corresponding to the value on energy axis) and the normalized magnetix flux $n$, and the colored indicators show the Bott index (\ref{BOTT}) obtained with the parameters $(\mu,n)$.
          To calculate the Bott index, we choose the chemical potential corresponding to the center of gap.
          The vertical dashed lines represent multiples of $N_\rr=843$ (red) and $N_\rb=521$ (blue).
	}
	\label{fig:HousdatoferAHM}
\end{figure}

With the configuration of Peierls phase $\{\vp_{ij}\}$, we obtain the energy spectrum by means of numerical diagonalization of the Hamiltonian (\ref{hamil}).
Figure~\ref{fig:HousdatoferAHM} shows the energy spectrum in a $g=6$ Penrose approximant.
Since the numbers of tiles $N_\rr$ and $N_\rb$  for given generation $g$ are  coprime, the periodicity for the normalized staggered flux $n$ is $N_\rr\times N_\rb$, corresponding to $[0,2\pi N_\rb]$ for $\phi_\rr$ and $[0,2\pi N_\rr]$ for $\phi_\rb$.
With increasing the generation to the thermodynamic limit, i.e., $N_\rb\to\infty$, the periodicity of energy spectrum with respect to the staggered magnetic flux $\phi_\rr$ or $\phi_\rb$ no longer remains.

\begin{figure}[t!]
	\centering
				\includegraphics[width=\linewidth]{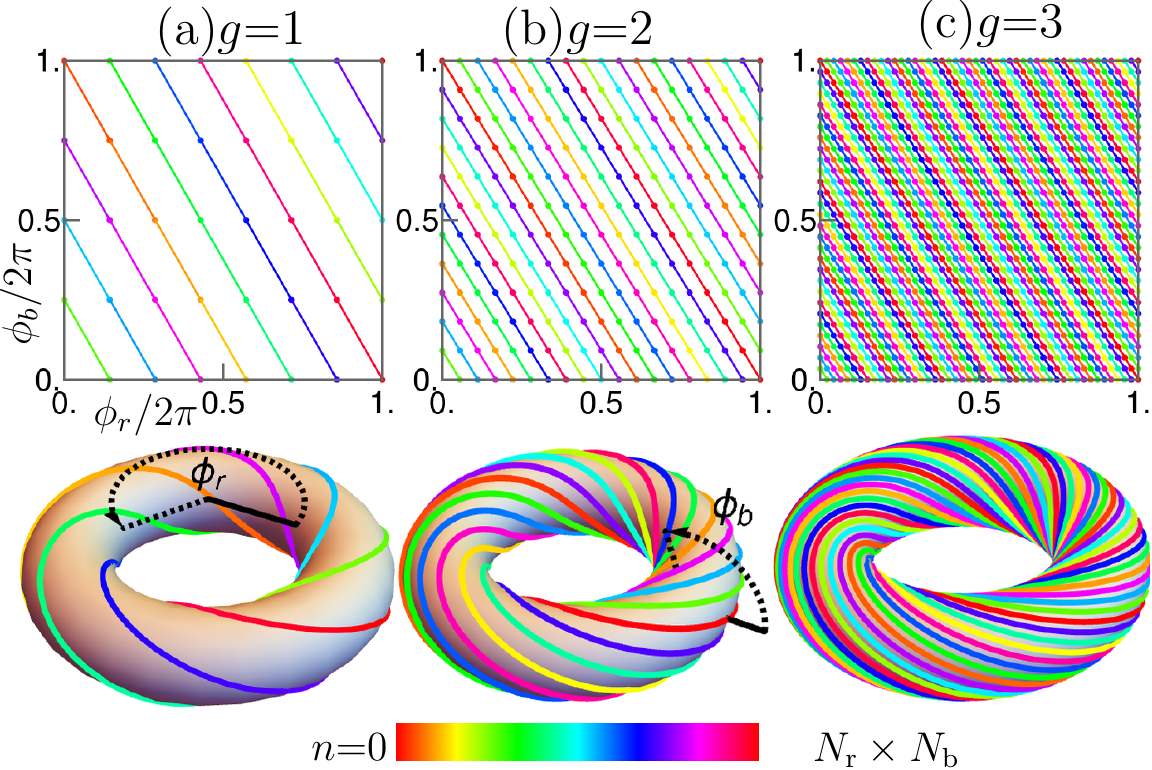}
	\caption{The path of possible region of the staggered magnetic fluxes $(\phi_\rr,\phi_\rb)$ for $g=1,2,$ and $3$ generations. Since the magnetic flux $\phi_\rr$ ($\phi_\rb$) is unique modulo $2\pi$, the parameter space of two flux corresponds to a 2-torus, where the toroidal (poloidal) axis denotes $\phi_\rr$ ($\phi_\rb$). With increasing the generation, the path densely covers the whole region of the parameter space, due to approaching the irrational ratio of the fluxes.
}
	\label{fig:trajectory}
\end{figure}

Next, we discuss the topological features by using the Bott index~\cite{Fulga2016,Ghadimi2021,Loring2019} defined by,
\begin{equation}\label{BOTT}
	B_\rt=\frac{1}{2\pi}\Im\tr\log \left(VUV^\dagger U^\dagger\right),
\end{equation}
with
\begin{equation}\label{XYUV}
	U=P \mathcal{X} P+(I-P),\ V=P \mathcal{Y} P+(I-P),
\end{equation}
where $P$ is the projection matrix onto lower-lying states than the Fermi level and $I$ is the identity matrix.
$\mathcal{X}$ and $\mathcal{Y}$ are the diagonal matrices linking the position to a U(1) phase,
\begin{align}\label{XPBC}
	\mathcal{X}_{i,j}&=\exp(2\pi \imath \frac{x_i-x_\mathrm{min}}{x_\mathrm{max}-x_\mathrm{min}}) \delta_{i,j}, \\
        \mathcal{Y}_{i,j}&=\exp(2\pi \imath \frac{y_i-y_\mathrm{min}}{y_\mathrm{max}-y_\mathrm{min}}) \delta_{i,j},
\end{align}
where $x_\mathrm{min}$ and $x_\mathrm{max}$  ($y_\mathrm{min}$ and $y_\mathrm{max}$) are minimum and maximum values of the $x$ ($y$) component of position, respectively, and $\delta_{i,j}$ is the Kronecker delta.
The Bott index gives non-zero integer values in nontrivial topological phases as in the case of other topological invariant.
In particular, the Bott index is useful for real-space representation of wavefunctions, and can be obtained with the periodic boundary condition.
Moreover, since the Bott index basically suits the model without the time-reversal symmetry, we choose it to calculate the topological number in our model, where the time-reversal symmetry is broken due to the staggered magnetic flux.
In Fig.~\ref{fig:HousdatoferAHM}, we can see some gaps including colored symbols.
The colored symbols represent the Bott index at the parameter point of the chemical potential and the normalized flux.
We have also checked the appearance of edge modes in the topological phases with the open boundary condition (not shown).

\begin{figure*}[t!]
	\centering
	\includegraphics[width=0.65\linewidth]{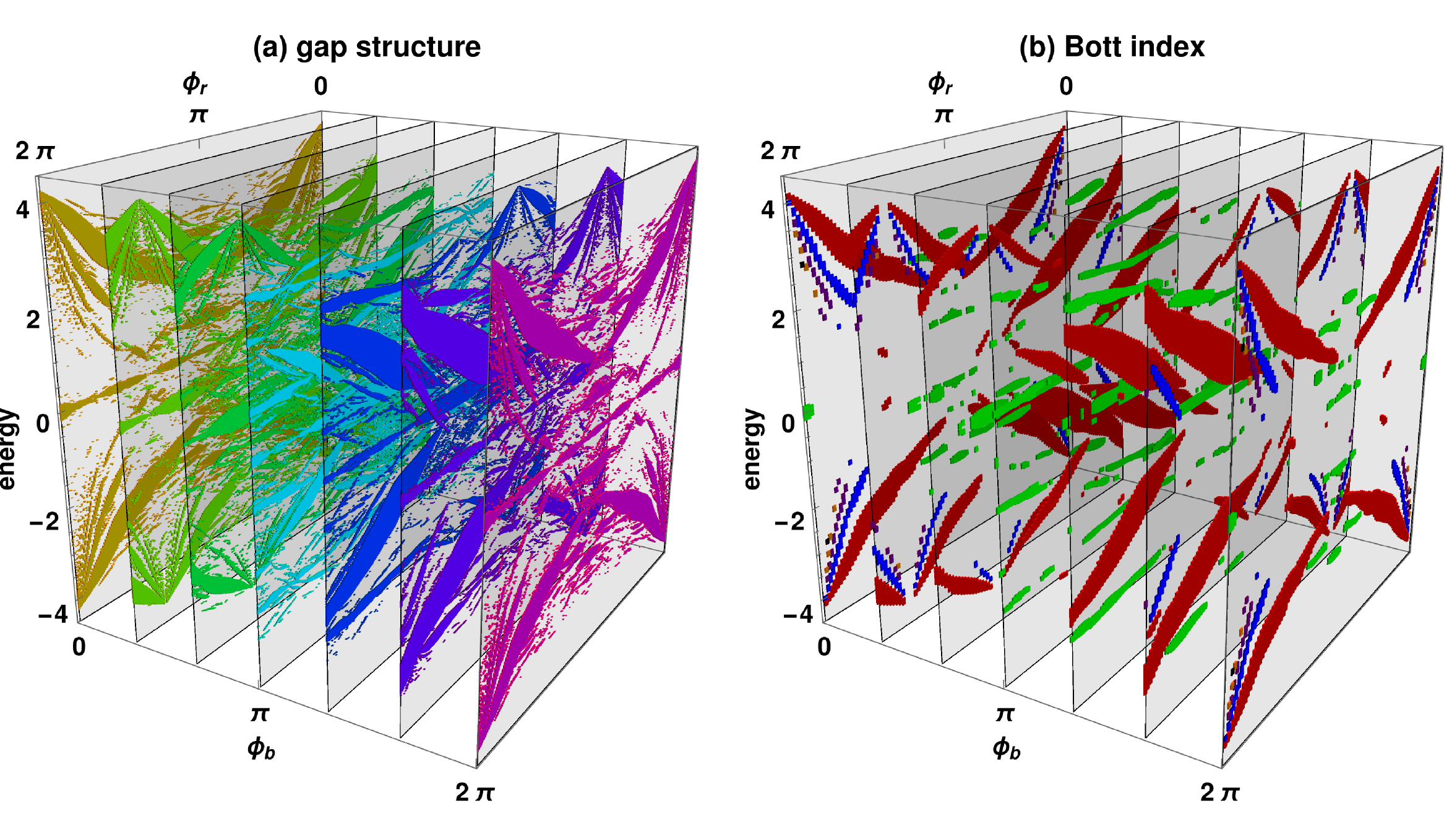}
	\caption{Three-dimensional Hofstadter-butterfly structure. (a) Gap structure and (b) Bott index in energy spectrum as a function of two independent magnetic fluxes $\phi_\rr$ and $\phi_\rb$ in a $g=6$ Penrose lattice.
		In (a), we show the gap position if the gap $\Delta E>0.01 t$. In (b), the Bott index is shown for the gaps with $\Delta E>0.05 t$.
		The colors in (b) are the same as Fig.~\ref{fig:HousdatoferAHM}.
	}
	\label{fig:HousdatoferFull}
\end{figure*}

As mentioned above, the magnetic flux is unique modulo $2\pi$, i.e., $\phi_\rr$ and $\phi_\rb$ can be folded into $[0,2\pi]\times[0,2\pi]$, corresponding to a 2-torus.
In Fig.~\ref{fig:trajectory}, we plot the path of flux for several generations, due to the constraint $\phi_\rb=-(N_\rr/N_\rb)\phi_\rr$.
Apparently, the possible region of $(\phi_\rr,\phi_\rb)$ is expanded with increasing the generation, because of the increase of the smallest common multiple of $N_\rr$ and $N_\rb$.
In the $g\to\infty$ limit, namely in the quasicrystals, the ratio converges to an irrational number (the golden ratio) $N_\rr/N_\rb\to\tau$ [see Table.~1], so that the possible region of $(\phi_\rr,\phi_\rb)$ densely covers the whole space of the 2-torus.
Consequently, in the quasiperiodic limit, the magnetic fluxes $\phi_\rr$ and $\phi_\rb$ are no longer dependent each other, and are regarded as two independent parameters.
Note that this feature is inherent in the quasicrystals, in contrast to periodic systems like kagome lattice~\cite{Xu2015}.

The independence of the magnetic fluxes $\phi_\rr$ and $\phi_\rb$ together with an energy axis implies the existence of a three-dimensional structure of energy spectrum.
Figure~\ref{fig:HousdatoferFull} shows the gap structure in the energy spectrum with two parameters $\phi_\rr$ and $\phi_\rb$.
We can see that the complicated structure, the so-called Hofstadter butterfly, spreads out in three dimension.
	With changing the staggered magnetic flux, system probes this two-dimensional flux phase space in a trajectory shown in  Fig.~\ref{fig:trajectory}.
Note that the length of this path in true quasicrystal limit ($g\to \infty$) is infinite, originating form the golden ratio in the Penrose lattice.

Finally, we again insist on the similarity and difference between our staggered magnetic flux model and a model with a uniform magnetic field in the quasicrystals proposed in \cite{Duncan2020}.
The latter model captures  an incommensurate behavior of magnetic flux due to the irrational ratio of surfaces of two tiles, and therefore the phase diagram is an aperiodic function of the magnetic field similar to our model.
Folding a resulting energy spectrum as a function in $(\phi_\rb,\phi_\rr)$ phase space, one can obtain the same three-dimensional Hofstadter butterfly as ours model.
However, as mentioned in \cite{Duncan2020}, one can find the irrational ratio of surfaces of two tiles even in a uniform lattice, e.g., it consists of two rectangles whose surfaces have an irrational ratio.
This point is an essential difference between the uniform and staggered fields.
In our model, with the staggered field, the aperiodicity of the phase diagram stems from the irrational ratio of the numbers of two tiles.
The ratio of the numbers of tiles can be irrational only in quasicrystals, and thus the present feature that we found is unique in quasicrystals.


	In conclusion, we have investigated a tight-binding spinless fermion system on the two-dimensional quasicrystals composed of two types of tiles, with an emergent staggered magnetic flux, without external magnetic fields.
We found that the energy spectrum and its topological properties are aperiodic as a function of staggered magnetic flux.
Furthermore, due to the irrationality of the ratio of the number of the two tiles, two local fluxes are eventually independent in quasicrystals.
Consequently, the flux phase space is two-dimensional, and increasing staggered magnetic flux probes this flux phase space in a nontrivial way, i.e., the way never overlap.
We found that the energy spectrum in this two-dimensional phase space produces a Hofstadter butterfly-like fractal structure, but lives in three dimensions.

\section{Acknowledgments}
R.G. was supported by the Institute for Basic Science in Korea (Grant No. IBS-R009-D1), and the National Research Foundation of Korea (NRF) grant funded by the Korea government (MSIT) (No.2021R1A2C4002773, and No. NRF-2021R1A5A1032996).
T.S. was supported by the Japan Society for the Promotion of Science, KAKENHI (Grant No. JP19H05821).

\bibliography{refs-b}
\end{document}